  \documentclass[multi]{cambridge6Atight}
 
  \usepackage{natbib}

  \usepackage{rotating}
  \usepackage{floatpag}
  \rotfloatpagestyle{empty}

\usepackage{amsmath}% if you are using this package,
  \usepackage{amsthm}
  \usepackage{graphicx}
  \usepackage{mathptmx}

  \usepackage{makeidx}
  \makeindex

 \copyrightline{Reprinted from \textit{The Impact of Binaries on Stellar Evolution}, Beccari G. \& Boffin H.M.J. (Eds.), \copyright\ 2018 Cambridge
    University Press.}
    
\begin{document}

%%%%%%%%%%%%%%%%%%%%%%%%%%
%%%% START TO EDIT FROM HERE %%%%%%
%%%%%%%%%%%%%%%%%%%%%%%%%%

  \alphafootnotes
   \author[F Patat and N Hallakoun]
    {Ferdinando Patat
    and Na'ama Hallakoun}
  %%%% PUT HERE TITLE OF YOUR CHAPTER  
  \chapter{Type Ia Supernovae: Where Are They Coming From and Where Will They Lead Us?}
  %%% Put A. Einstein et al. in [] if more than 3 authors - these are running authors
  %%% \chapter[Einstein et al.]{Writing a chapter for the Imbase17 book}

%%% footnotes are not compulsory... please use only if needed. Don't put your affiliation here.
  %\footnotetext[1]{fpatat@eso.org}
  \arabicfootnotes

%%% Provide here again for all authors of the chapter, the name and affiliation -- only one author at the time!
  \contributor{Ferdinando Patat
    \affiliation{European Souther Observatory,
    Karl-Schwarzschild-str. 2, 85748 Garching, Germany}}
    
\contributor{Na'ama Hallakoun
    \affiliation{School of Physics and Astronomy, Tel-Aviv University, Tel-Aviv 6997801, Israel}}

%%%% PUT HERE YOUR ABSTRACT
 \begin{abstract}
We present a summary of our understanding of Type Ia Supernova progenitors, mostly discussing the observational approach. The main goal of this review is to provide the non-specialist with a sufficiently comprehensive view of where we stand.
 \end{abstract}

\section{Source of Embarrassment vs. Excitement}
\label{PatatSec1}
%%%% Replace Author by your last name please %%%%
%%% Sec1 can be replaced by what you prefer %%%%
About ten years ago we opened a similar review \citep{patat09} with the following quotation from Mario Livio: 'The fact that we do not know 
yet what are the progenitor systems of some of the most dramatic explosions in the universe has become a major embarrassment 
and one of the key unresolved problems in stellar evolution' \citep{livio}. Today, almost twenty years after Livio had expressed his 
feelings on the matter, and although we have arguably learned a lot on this topic, there still are good reasons for being embarrassed. 
Or, if you are an enthusiast, to keep the excitement high.

In this review, which is explicitly meant for non-Supernova specialists, we will give a review of our understanding of Type Ia progenitors. 
For more details the reader is referred to the exhaustive review by \cite{Maoz_2014}.

\subsection{Why and How?}
\label{PatatSubsec1}

One may think that, once having established that Type Ia Supernovae (hereafter SN Ia) are good standard candles, there is no need for understanding what they actually are. After all, a candle is a candle is a candle: why should it be important to know what wax it is made of?

We reckon there are a number of reasons, which can be summarised as follows:

\begin{itemize}
\item Using SNe Ia in Cosmology requires an understanding of the evolution of the luminosity, and the SN rate with cosmic epoch. Both depend on the nature of the progenitors;
\item Galaxy evolution depends on the energetic and nucleosynthetic output of SNe Ia;
\item A knowledge of the initial conditions and of matter distribution in the environment of the exploding star is essential for the understanding of the explosion itself;
\item An unambiguous identification of the progenitors, coupled with observationally determined SN rates will help placing constraints on the theory of binary star evolution.

When addressing the progenitor problem problem, there are a few fundamental facts that need to be kept in mind:

\begin{enumerate}
\item SNe Ia do not show Hydrogen lines;
\item They are homogene-ous[isable] in terms of peak luminosity, spectra and light curve shape;
\item They appear in elliptical galaxies (as opposed to core-collapse SNe);
\item The above rule out core-collapse of massive ($M>8 M_\odot$), young stars.
\end{enumerate}
\end{itemize}

From this comes the conundrum that has been bothering the SN community for a long time: a) you need to make a low-mass, 
`old' star explode; b) you need to produce
almost always about the same luminosity (i.e. the same amount of radio-active $^{56}$Ni), and c) you ought to hide the most 
abundant element in the universe (which, in itself, is quite a remarkable fact).

The theoretical solution to this riddle was first put forward by \cite{wi73}, who invoked 'mass transfer between members of a binary system'.

\subsection{The Basics}
\label{PatatSubsec2}

Well before the above scenario was proposed, from the observations of both SNe and SN remnants it was clear 
that energetics and composition require the thermonuclear combustion of a degenerate object \citep{hf60}. In addition,
the absence of H and He indicates that the exploding object is most likely a C-O White Dwarf (for which the Chandrasekhar mass
is $M_{CH}\sim$1.4 M$_\odot$).

The general understanding is that the core of the star is burned into Fe-group elements plus some Intermediate Mass Elements (IME). 
The observed luminosity (typically peaking at $M_V\sim-19.3$) requires $\sim$0.6 M$_\odot$ of radio-active $^{56}$Ni.

In terms of energetics, this can be illustrated with a few back-of-the envelope calculations. The nuclear energy of 0.6 $M_\odot$ of Ni
is about 10$^{51}$ erg. A similar amount is derived for the remaining 0.8 $M_\odot$ of Iron-peak elements, so that the total nuclear energy
is $E_N\sim$2$\times$10$^{51}$ erg. Of this, about 1\% is radiated as visible light ($\sim$10$^{49}$ erg).
The gravitational binding energy of a  1.4 M$_\odot$ C-O WD is $E_G\sim$0.5$\times$10$^{51}$ erg. Therefore $E_N>E_G$, so that the 
thermonuclear explosion is indeed capable of completely disrupting the WD. Given that only a small fraction of the total energy is
released in the form of radiation, we can estimate the expected ejecta velocity assuming that all energy is converted to kinetic energy: $v_e\approx\sqrt{2 E_K/M_{WD}}\sim$10$^{4}$ km s$^{-1}$, which is indeed what is observed. Therefore, the hypothesis holds in terms of energy budget.

All this led the community to the following consensus statement: SNe Ia represent thermonuclear disruptions of mass accreting C-O WDs, when these approach the Chandrasekhar limit and ignite carbon. 

Somehow, sometime and somewhere.

As it turns out, growing a WD to the critical limit is not that easy: if our understanding of stellar evolution is correct, there is no way a single star can produce a WD with a mass close to $M_{CH}$. Therefore, one way or another, the WD needs to accrete material, either from a non-degenerate companion (via some mass transfer mechanism, like Roche-lobe overflow or wind) or a degenerate companion (via merging or collision). These two scenarios are commonly dubbed single-degenerate (SD) and double-degenerate (DD).

\section{The SNe Ia Progenitor Problem: an Observational Approach}
\label{PatatSec3}

More than forty years have passed from the first suggestion for solving the conundrum and, despite SNe Ia are used for precision cosmology, the nature of the progenitor system[s] is still unknown.

Very schematically, the problem has been addressed through the following channels:

\begin{itemize}
\item Populations of potential progenitors;
\item Pre-explosion imaging of [nearby] explosion sites;
\item Explosion properties carrying progenitor's imprints;
\item Circum-stellar environment;
\item SN remnants;
\item Explosion rates as function of time and location;
\item Binary population synthesis.
\end{itemize}

We will briefly discuss each of them in the next sub-sections.

\subsection{Candidate Populations}
\label{PatatSubSec21}

One thing is speculating about the existence of binary systems with $M\sim M_{CH}$ accreting WD, another is looking around for real examples and see what they tell us. Among the SD scenario, the following binary systems have been considered as possible candidates: Recurrent Novae (RNe), Super-Soft X-ray Sources (SSS), Rapidly Accreting WDs and He-rich donors. On the DD side, the attention has obviously focused on binary WD systems.

A fair amount of work went into investigating RN systems like U Sco, RS Oph, T CrB, all containing WDs claimed to have masses close to the critical limit. However, a number of questions remain open. For instance: is the WD really a C-O WD? If this is instead a O-Ne-Mg WD, then the star would not explode as a Type Ia. In these systems it is not possible to directly observe the properties of the WD (for instance because it is embedded in an accretion disk), and so this is a question likely doomed to remain unanswered. Another crucial point is related to the accretion: these systems may host a massive WD, but one cannot be sure its mass is increasing with time. There is the possibility that during the RN phase, the WD actually loses at each outburst more mass than it accretes during the inter-outburst phase. Finally, there is a fundamental issue about the number of these systems in the Milky Way (MW), which may well not be sufficient to account for the observed rate.

In general, SD systems are supposed to spend some Myr in a phase of steady nuclear burning while accreting material from the donor. In virtue of this fact they should be detectable as super-soft X-ray sources \citep{distefano2010}. Under certain conditions, also DD systems may undergo  a SSS phase, with much lower luminosities, though.
Therefore, the detection of a SSS in coincidence with a SN Ia explosion site would allow a direct study of at least some of the progenitor's properties. This has been first applied to SN2007on, for which a pre-explosion detection, based on X-ray imaging was claimed, but turned out not to be fully convincing \citep{voss08,roelofs08}. The conclusion is that if the bulk of Type Ia is to be explained by SSS, only 1\% of the WD mass accretion takes place in this phase to match the observed MW rate \citep{Maoz_2014}.

Until a few years ago, binary WD systems, first suggested by \cite{tutukov81},  were considered as a disfavoured scenario. First of all because it was not clear whether there were enough suitable systems in nature, and second because the range of WD masses was seen as being in contradiction with the observed homogeneity of SNe Ia. In addition, from a theoretical point of view, the WD merger was suspected to result into an Accretion Induced Core Collapse (AICC), leading to the formation of a neutron star rather than to a thermonuclear explosions \citep{nomoto85}.

The first large-scale attempt to find suitable WD-WD candidates was the ESO-SPY search \citep{Napiwotzki_2004}. This campaign included about one thousand WDs and led to the detection of one system with a global mass larger than $M_{CH}$ and that would merge in less than one Hubble time. Other similar studies are presented by \citet{badenes09} and \cite{Maoz_2017}. Of course, like for the SD candidates, the crucial question is whether there are enough such systems to sustain the observed rate. We will come back on this in greater detail in Section~\ref{PatatSec5}.

\subsection{Pre-Explosion Sites}
\label{PatatSubSec22}

The idea is very simple: one determines the SN position as accurately as possible and goes back to pre-explosion archival data and searches for traces of the precursor system. While this had led to a number of positive detections for core-collapse systems, this has never been the case for thermonuclear SNe and only not really stringent constraints on the progenitor's luminosity could be placed. Until the nearby SN~2011fe exploded in M101. This was a very close-by ($\sim$6 Mpc) and very early (discovered most likely a few hours after explosion) standard Type Ia, which provided a unique opportunity to probe the earliest phases in great detail, across a wide wavelength range. Being the host a Messier object, rich pre-explosion, HST data were available and this allowed Weidong Li and collaborators to draw some firm conclusions \citep{li2011}: a red giant at the location of the SN is ruled out in the decade before the explosion, hence removing RS-Oph and T CrB systems as viable progenitors. Stars with $M>3.5M_\odot$ are also excluded, while a main sequence or sub-giant donor is allowed (i.e. systems like U Sco in quiescence). A red giant was also ruled out for another, even closer (3.5 Mpc) event: SN~2014J \citep{kelly14}.

\subsection{An Interlude on WD Spin-Up/Down}
\label{PatatSubSec23}

The common understanding of the physics of SNe Ia is that the exploding star has to reach some critical condition while approaching $M_{CH}$. Nevertheless, WD rotation can have a significant influence, as it has been shown in the so called spin-up/spin-down scenario in the context of SD progenitors: a WD that has grown in mass, even beyond the critical limit, could be rotation-supported against collapse and ignition during the accretion. This was first introduced by  \cite{yoon04}, and it has received quite some interest in recent years. If this mechanism is at work, the traces of the process (including the donor) could be completely wiped out by the time of the explosion. Interestingly, DD mergers explosions delayed by rotational support have also been proposed \citep{amedeo14}.

The bottom line is that there may be a population of rapidly spinning WDs, waiting to spin-down and eventually explode. This is of course a rather frustrating thought, very effectively phrased by \cite{Maoz_2014}: `Observationally, a spin-up/spin-down scenario could potentially 'erase' many of the clues we discuss in this review'.

\subsection{Explosion Properties}
\label{PatatSubSec24}

A number of aspects seen during the explosion relate (unfortunately often ambiguously), to the progenitor properties:

\begin{itemize}
\item Early light curve and spectral evolution;
\item Shocks on [possible] companion star;
\item Ejecta asymmetries (polarimetry);
\item Emission from hydrogen;
\item Radio, X-Ray CSM emission;
\item Narrow time-variable absorptions;
\item CS Dust and light echoes (spectrophotometry and polarimetry).
\end{itemize}

Each of these channels has been explored in depth. Here we will focus on the first two, as these were fundamental for studying the case of SN~2011fe, which provided the most stringent constraints ever. The rationale is very simple: in the phases immediately following the thermonuclear runaway, the interaction between the SN ejecta and a stellar object in the vicinity of the explosion is expected to produce some effects, which depend, for instance, on the size of the star and the extent of its atmosphere \citep{kasen2010}.

For SN~2011fe, \cite{nugent11}, based on very early optical and UV observations, were able to exclude the presence of shock effects from SN ejecta hitting an alleged companion, definitely ruling out a RG companion, in line with the independent conclusions reached by \cite{li2011}. \cite{bloom2012} went one step further, ruling out even a main-sequence (MS) star, hence concluding that the DD scenario is the most probable for this one object. In addition, for the first time, their observations provided a direct upper limit to the size of the exploding object, $R<0.02 R_\odot$, so that the exploding star can be either a WD or a neutron star.
Ongoing and planned surveys have the potential of leading to much more numerous very early discoveries, so that we believe this channel of study will gain importance in the future.

The study of overall asymmetries in the ejecta has also been used to set constraints on the progenitor. In general, the fact that normal Type Ia show a continuum polarization consistent with zero has normally been interpreted as a sign of spherical symmetry \citep[see for instance][]{patat09b} which, in turn, is probably inconsistent with explosion scenarios that imply a high degree of asymmetry, like the case of violent mergers or head-on collisions \citep{bulla16}. Interestingly, different polarization properties were derived for sub-luminous events, possibly pointing to an explosion arising within a fast rotating WD \citep{patat10}.

So far no emission coming from the interaction between SN ejecta and circumstellar material has been detected, hence indicating a clean environment, in turn favouring a DD scenario. At least to a first glance. Spin-up/down can of course delay the explosion so much that the material lost by the progenitor system before the explosion has a sufficient time to diffuse into the inter-stellar medium. This would make that material undetectable, at least through emission (optical or X-ray). But it may be detectable in absorption, using trace elements like Na or K, which enable the observability of very tiny column densities. In addition, if the gas is far enough to be 'untouched' by the ejecta, but close enough to be affected by the SN radiation field, one may be able to observe time dependencies. This was the idea behind a series of studies following \cite{patat07}, who detected a marked variability in the well studied SN~2006X. As the variability is expected to take place only for a limited range of distances and physical conditions, hence taking very long to accumulate a sufficient sample, other searches have moved towards a more statistical approach, looking for an excess of blue-shifted Na absorption features signalling the presence of material lost by the progenitor \citep{assaf11,kate13}. These studies led to the conclusion that $\sim$20\% of Type Ia show these features. Unfortunately the interpretation is not univocal in terms of the progenitor's nature, as this gas may both be material lost by the non-degenerate companion, but also the relic of the common-envelope phase of a DD system.
 
We finally mention the polarization studies along the lines of sight to reddened Type Ia, as they represent a recent addition to the toolbox. Very intriguingly, the interstellar polarization derived for these objects, shows a polarization dependence that diverges in a remarkable way from what is seen in our own Galaxy \citep{patat15}. Excluding the anti-copernican conclusion that there is something special with the MW dust, the alternative explanations rely on the special properties of the explosion environment. A further study has shown that, as a matter of fact, it is not true that there are no stars in the Galaxy showing these peculiar properties. In fact, very similar polarization laws are observed in a sub-set of objects embedded in proto-planetary nebulae \citep{aleks17}.

Although there still are some unclear aspects, this may provide some support to the alternative core-degenerate (CD) scenario proposed by \cite{kashi11}. In this model, the WD merges with an asymptotic giant branch (AGB) star during the common envelope phase. The envelope is ejected, the explosion of the merged core is delayed by rotation, it spins down (via magnetic dipole radiation) and eventually explodes. This model, which broadly belongs to the DD family, has the advantage of clearing the immediate surroundings, and still leave material to be seen in absorption, which would reconcile it with the results reported above.

\subsection{Surviving Companions}
\label{PatatSubSec25}

Obviously, one substantial difference between DDs and SDs is the presence of a companion that becomes unbound after the explosion, and it hence runs away from the explosion site. The models predict that the companion star survives the explosion, it gains an anomalous transversal and rotational speed, and it probably gains a peculiar chemical composition in the atmospheric layers, caused by the pollution by SN ejecta.
With this concept in mind (no star $\Longleftrightarrow$ no SD), a few teams set out with the following agenda: search historical SN remnants (SNRs) for weird composition, fast moving, rapidly rotating stars not far from the estimated explosion centre.
A positive detection was first claimed by \cite{pilar04} for the so-called star G in Tycho's remnant. This was not confirmed by \cite{wolf09, wolf13} \citep[but see the rebuttal by][]{bedin14}.
No detection was reported for SN~1006 \citep{wolf12}, for SNR0509-67.5 \citep{schaefer12} and for Kepler \citep{wolf14}.
We believe it is fair to state that, so far, there is no convincing evidence for direct companion detection.

\subsection{SN Remnants}
\label{PatatSubSec26}

The idea is to look for possible interactions between the remnant and pre-explosion material, and to compare with the predictions of hydrodynamical modelling. In the SD scenario, one expects a cavity to be blown in the ISM (3-30 pc) during the fast-wind phase of the accretion \citep{hachisu96}. Such cavity was not detected in 7 Type Ia SNR \citep{badenes07}. In addition, X-ray observations are consistent with a uniform ISM density. Therefore, either the fast-wind phase does not take place, or the accretion stopped well before the explosion.

\subsection{SN Rates and Binary Population Synthesis}
\label{PatatSubSec27}

The Galactic rate of Type Ia explosion is known: $\sim$5$\times$10$^{-3}$ yr$^{-1}$. This can be used to constrain the progenitor scenarios using Binary Population Synthesis (BPS). 
The models show that, in principle, DDs can produce this value, while SDs can only explain some fraction of the observed rate, with WD+MS being the most productive, and WD+RG the less productive.
The observed rate (as a function of time) results from the combination of the star formation rate (SFR) and the so-called delay time distribution (DTD), which provides the time between the birth of the progenitor system and the explosion. Using different DTDs produced by BPS, can be directly compared to observed rates as a function of redshift. The match provided by DDs is definitely better than that deduced for SDs 
\citep[see][for a comprehensive discussion]{Maoz_2014}.

\vspace{5mm}

From the above it is clear that, although there still is some room for the SD channel, the DD scenario has gained quite some interest, as it may indeed explain the bulk of Type Ia. For this reason in the next Section we discuss in some detail its viability.

\section{Are There Enough Double White Dwarf Mergers to Explain the Milky Way's SN Ia Rate?}
\label{PatatSec5}

If most Type-Ia supernovae\index{type-Ia supernova} (SNe Ia) share the same formation channel, then any candidate scenario competing for the title should have enough progenitor systems to account for the observed SN Ia rate in the Galaxy. The situation of the SD\index{single degenerate} scenario in this sense is rather poor: even the most optimistic estimates for the number of recurrent novae\index{recurrent nova} in the Milky Way\index{Milky Way} \citep{Pagnotta_2014}, cannot account for the majority of SN Ia explosions. But are there enough double white dwarf\index{white dwarf!double} (WD) systems to explain the Galactic SN Ia rate through the DD channel? In order to answer this question we need to characterise the local double WD population.

The first systematic searches for detached double WDs were carried out in the 1980s, mostly with no detections \citep{Robinson_1987, Foss_1991}, or with a low rate of success \citep{Bragaglia_1990}. \citet{Marsh_1995} decided upon a different approach, and focused their attempts on low-mass ($<0.45$\,M$_\odot$) WDs, that have not had enough time to evolve from single stars during the lifetime of the Universe. Their method proved fruitful, with five out of seven cases turning out to be double WDs. However, the eventual mergers of these low-mass systems are not likely to result in a SN Ia explosion. Further studying a sample of 46 WDs, \citet{Maxted_1999} estimated a binary fraction of $1.7-19\%$. Combined with a few other surveys that were carried out in the 1990s, 18 close double WDs were found among a total of $\sim 200$ WDs surveyed \citep[see][and references therein]{Napiwotzki_2004}. An important step forward was initiated in the early 2000s, with the ESO SN Ia Progenitor surveY\index{SPY} \citep[SPY;][]{Napiwotzki_2001}, that took advantage of the then-new ESO 8\,m Very Large Telescope (VLT), to acquire unprecedented high-resolution, high signal-to-noise (S/N), multi-epoch spectroscopy of about 800 bright WDs. A number of newly discovered double WDs from the SPY survey were then followed-up and published \citep[e.g.][]{Napiwotzki_2002, Karl_2003a, Karl_2003b, Nelemans_2005}. To date, there are about 100 close double WD systems with known orbital parameters, including seven eclipsing double WDs \citep[see][and references therein]{Brown_2017}. However, when we exclude systems containing extremely low-mass (ELM) WDs\index{white dwarf!ELM} with $M < 0.3$\,M$_\odot$ \citep[e.g.][]{Brown_2016}, this number is reduced to $\sim 30$.

To estimate the WD binary fraction and its merger rate using a well-defined statistical approach, \citet{Maoz_2012} devised a method that characterises the local WD population by measuring the distribution of $\Delta$RV$_{\rm max}$, the maximal radial-velocity\index{radial velocity} (RV) shift between different epochs of the same WD. They showed that even with noisy and sparsely-sampled data (with as little as two epochs per system), the $\Delta$RV$_{\rm max}$ distribution can provide meaningful constraints on the WD binary fraction and on the binary separation distribution, as well as on the double WD merger rate. The `core' of the $\Delta$RV$_{\rm max}$ distribution, at low RV differences where most of the systems reside, is the result of the measurement uncertainty, and WDs that are found in this area of the distribution are most likely single. However, systems that end up in the `tail' of the distribution are most probably real detections of double WDs. The observed $\Delta$RV$_{\rm max}$ distribution is then compared to simulated distributions with varying binary fraction and separation distribution parameters: populations with a high binary fraction or a separation distribution favouring systems at closer separations introduce a higher fraction of `tail' systems. Thus, this method allows probing of the double WD population characteristics without the need of follow-up observations and full orbital solutions of the binary candidates.

The $\Delta$RV$_{\rm max}$ method requires the use of a large spectroscopic sample, with at least two epochs per WD. Exploiting the multiple sub-exposures per object taken by the Sloan Digital Sky Survey (SDSS)\index{SDSS} to allow cosmic ray rejection, \citet{Badenes_2012} applied this method to a large sample of $\sim 4000$ WDs that have SDSS spectroscopic data. They found that the double WD merger rate is of the order of magnitude of the SN Ia rate in the Milky Way (based on the rate measured for Sbc galaxies and the approximate Galactic mass). However, because of the large measurement errors of the SDSS spectra ($\sim 80$\,km\,s$^{-1}$), only a few, high-orbital-velocity, candidate systems could be detected with small-number statistics, and the measured merger rate was highly uncertain. The low RV precision was mostly the result of the low resolution of the SDSS spectra, combined with the fact that most WDs show only very broad hydrogen Balmer absorption lines in their optical spectra. Thus, only double WD systems with $\Delta$RV$_{\rm max}$ higher than $\sim 250$\,km\,s$^{-1}$ (or with orbital separations smaller than $\sim 0.05$\,AU), constituting 15 out of the $\sim 4000$ WDs, were detectable in the SDSS sample.

In order to improve the results and remove the degeneracy between the model parameters, a more precise multi-epoch spectroscopic sample was needed -- a requirement fulfilled by the above-mentioned SPY\index{SPY} sample \citep{Napiwotzki_2001}. Owing to the high resolution of the SPY sample, finer details in the WD spectra could be used to extract the RV with a precision of $\sim 1-2$\,km\,s$^{-1}$ per epoch. Applying the $\Delta$RV$_{\rm max}$ method to a sample of 439 WDs from SPY, \citet{Maoz_2017} were able to further constrain the WD binary fraction and separation distribution, this time to separations out to $\sim 4$\,AU, with 43 candidate double WD systems with $\Delta$RV$_{\rm max}>10$\,km\,s$^{-1}$. Since the sample was smaller by an order of magnitude compared to the SDSS sample of \citet{Badenes_2012}, it was not sensitive to rare systems with very small separations ($<0.05$\,AU), and thus the two samples were in fact complementary. In a recent study, \citet{Maoz_2018} combine the SDSS and SPY results to obtain improved constraints on the double WD population and its merger rate. They find that the WD binary fraction out to 4\,AU is about $10\,\%$, and that the Galactic specific WD merger rate is $\sim 10^{-11}$\,yr$^{-1}$ per WD. This rate is $4.5-7$ times higher than the Milky Way's specific SN Ia rate, implying that there are indeed more than enough double WDs to account for the observed SN Ia rate.

These results have further implications for the WD mass function. As \citet{Maoz_2018} note, when the double WD merger rate is integrated over the age of the Galaxy, it implies that $8.5-11\,\%$ of all WDs ever formed have already merged with another WD. Assuming that most of the mergers result in a more-massive WD rather than a SN Ia explosion, we expect that about $10\,\%$ of the WDs are double WD merger products. This is consistent with the $15-20\,\%$ ``high-mass bump'' at $\sim 1$\,M$_\odot$, observed in the WD mass function when including faint WDs \citep[see][and references therein]{Maoz_2018}.

These numbers suggest that $\sim12-18\,\%$ of the double WD mergers result in a SN Ia explosion. However, if a core density of at least $\sim 2 \times 10^7$\,g\,cm$^{-3}$ is needed for the production of $\sim 0.5$\,M$_\odot$ of $^{56}$Ni observed in SNe Ia, then the merging binary should consist of at least one WD more massive than $\sim 0.9$\,M$_\odot$ \citep[B. Katz, private communication;][]{Moll_2014}. This hints at a sequence of two mergers: one merger to create the massive WD, that later on merges with a third WD and explodes as a SN Ia. As \citet{Maoz_2018} note, this requires that nearly all of the merged double WDs have originated in triple systems. This is in contrast with the observed fraction ($10-20\,\%$) of triple systems among main-sequence A-type stars \citep{Duchene_2013, Leigh_2013}, which are the progenitors of most present-day WDs. This discrepancy might be resolved if the Galactic SN Ia rate is overestimated, or if the WD number density is underestimated. A better estimate of the WD number density is expected in the near future, with the results of the \textit{Gaia} mission. Nevertheless, the progenitor-population situation looks brighter for the DD scenario, with only a factor of a few discrepancy, compared to the more severe situation for the SD progenitor population.

\section{Single Degenerates: Is This It?}

In the last twenty years the paradigm has changed quite substantially, both from the observational (with the advent of massive SN searches) and the theoretical (with the deployment of 3D explosion simulations) point of view. In spite of the initial success (and popularity) of SDs, the DD scenario has become much more physically viable. 

All these new acquisitions have modified our perception, so much that the preference has moved from SD to DDs. At the 2013 Leiden workshop on Type Ia progenitors, we run a poll, asking the participants to give percentages (adding up to 100\%) for the various progenitor channels. A total of 47 valid ballots were received, and these are the results: SD: 22\%, DD: 60\%, CD: 7\%, Other: 11\%. Considering that CDs are a flavour of DDs, the popularity of double degenerates emerges very clearly. Of course, not all problems are solved. For instance, C-O + C-O mergers are likely too rare; this therefore requires sub-Chandrasekhar or C-O + He WD mergers. It remains to be seen whether this can explain the bulk of normal Type Ia in terms of observed properties (and not just their rates).

A few days before the start of the ImBaSE~2017 conference, we asked a number of expert opinions. We report them here, as they provide a nice complement to what we have tried to put together in this certainly incomplete summary.

\begin{enumerate}
\item  A review of all the proposed models reveals that each one of them still encounters a few significant difficulties. Consequently, the inescapable conclusion may be that Type Ia supernovae can be produced by a number of progenitor systems (Mario Livio).
\item The thermonuclear explosion of a C+O white dwarf has successfully explained the basic observed features of Type Ia supernovae. Both the Chandrasekhar and the sub-Chandrasekhar mass models have been examined. However, no clear observational indication rejects how the white dwarf mass grows until C ignition, i.e., whether the white dwarf accretes H/He-rich matter from its binary companion [SD] or whether two C+O white dwarfs merge [DD] (Kenichi Nomoto).
\item All men are created equal - but not type Ia supernovae. Nature does it right! (Wolfgang Hillebrandt)
\item A SN Ia is the outcome of detonating 1 solar mass of C and O with $\rho_{max}\approx$0.5--2$\times$10$^8$ g cm$^{-3}$ (Stan Woosley).
\item I recognise the need to consider a variety of models that might apply to the broad category of `Type Ia' and the limitations of the SD model, but still have reservations about DD models and sub-Chandra models (low central densities). I still think a delayed-detonation model, for its flaws, is the standard to beat in terms of reproducing the spectral evolution. Igniting carbon near the  Chandrasekhar mass seems to work best for typical Type Ia, but there are variations that might work for both SD and DD. I'm intrigued by spin-up/spin-down models (Craig Wheeler).
\item Just as all roads lead to Rome, and one size does not fit all, I think it's becoming increasingly clear that Type Ia supernovae are produced by several evolutionary paths leading to different progenitor systems and white dwarf masses (Alex Filippenko).
\item Do not be a slave to fashion!  Scientific ideas ebb and flow.  Though evidence for interaction is sparse and strong evidence for partners is missing, keep an open mind.  The single degenerate model is wounded, but not dead.  Perhaps nature finds more than one way to explode a white dwarf (Robert Kirshner).
\item Probably more than one way to make a SN Ia, but seems the primary track involves a white dwarf + another compact star (Brian Schmidt).
\item  At least two types of progenitor systems can produce SNe-Ia:  SD and/or DD. When the WDs reach the Chandrasekhar mass, they can explode in different ways: detonation, deflagration, delayed detonation... There are some occurrences in which WDs explode as sub-Chandra or super-Chandra. Therefore, from a phenomenological point of view, we potentially have a large variety of outcomes.  To some extent,  this fact seems supported by observation.  In conclusion there are different progenitors, different lifetimes, different sizes, different ages, different chemical compositions, and probably different spins. Nevertheless this variety is characterised by a common ending: after the explosion nothing is left:  `Much ado about nothing' or shall we think again? (Massimo Della Valle)
\item  Binaries in all cases, white dwarf in any case, leftover companion in no case (seen so far). Case to be made for binary evolution to produce the massive white dwarfs (Bruno Leibundgut).
\item As we are hesitant to choose between SD and DD, so may have been Nature, perhaps making prompt and late Ia's, respectively. Few, but not so few, intermediate mass binaries end up with a spectacular Type Ia display, as few of the more massive ones make black hole mergers and solar masses in gravitational waves. For us, understanding these paths is understanding how binaries evolve through two common envelope phases, which makes our job quite difficult, as uncertainties multiply uncertainties. But don't forget, haemoglobin comes from Ia's as chlorophyll from core-collapse supernovae. We have a special link to Ia's, hence an obligation to understand how they come about (Alvio Renzini).
\item In one sentence: Beware assumptions! Adding another:  Pre-and post explosion signatures of DD and SD models can be very much alike (Rosanne Di Stefano).
\item For typical SN Ia that eject about a Chandra mass, I like the canonical SD model (perhaps a DD model in which the merger product spins down long enough to become rather like the SD model, although with a total mass at least a bit above Chandra rather than a bit beneath it, might be OK), but if some SN Ia eject substantially less than a Chandra mass yet are basically symmetric, then the DD model with a Shen-Moore very low-mass helium shell may be best for them (David Branch).
\item This matter can be addressed by looking at what is the major challenge for each of the two scenarios, SD and DD, based solely on the MW SN Ia rate and on what we know about the putative progenitor populations (Dani Maoz).
\end{enumerate}
  
\bibliography{patat_bibl}\label{refs}
\bibliographystyle{cambridgeauthordate}
  
 %%%%%%%%%%%%%%%%%%%%%%%%%%% 
 \copyrightline{} 
 \printindex
    %this is to check if you are happy with your index -- it will not appear here in the book and you can ignore the additional page
    
%%%%%%%%%% END %%%%%%
\end{document}